\documentclass[aps,prep,epsfile,nofootinbib]{revtex4}

\usepackage{graphics}
\usepackage{slashed}
\usepackage{amssymb}
\usepackage{lscape}
\usepackage{amsmath}
\usepackage{graphicx}
\usepackage[dvipsnames]{xcolor}
\graphicspath{ {images/} }
\begin{document}
\title{Quantum thermodynamics in the interior of a Reissner-Nordstr\"om black-hole}
\author{$^{2}$ Juan Ignacio Musmarra\footnote{jmusmarra@mdp.edu.ar}, $^{1,2}$ Mauricio Bellini.
\footnote{{\bf Corresponding author}: mbellini@mdp.edu.ar} }
\address{$^1$ Departamento de F\'isica, Facultad de Ciencias Exactas y
Naturales, Universidad Nacional de Mar del Plata, Funes 3350, C.P.
7600, Mar del Plata, Argentina.\\
$^2$ Instituto de Investigaciones F\'{\i}sicas de Mar del Plata (IFIMAR), \\
Consejo Nacional de Investigaciones Cient\'ificas y T\'ecnicas
(CONICET), Mar del Plata, Argentina.}

\begin{abstract}
We study the interior of a Reissner-Nordstr\"om Black-Hole (RNBH) using Relativistic Quantum Geometry, which was introduced in some previous works. We found discrete energy levels for a scalar field from a polynomial condition for the Heun Confluent functions expanded around the effective causal radius $r_*$. From the solutions it is obtained that the uncertainty principle is valid for each energy level of space-time, in the form: $E_n\, r_{*,n}=\hbar/2$, and the charged mass is discretized and distributed in a finite number of states. The classical RNBH entropy is recovered as the limit case where the number of states is very large, and the RNBH quantum temperature depends on the number of states in the interior of the RNBH. This temperature, depending of the number of states of the RNBH, is related with the Bekeinstein-Hawking (BH) temperature: $T_{BH} \leq T_{N} < 2\,T_{BH}$.
\end{abstract}
\maketitle

\section{Introduction and motivation}\label{1}

The Reissner-Nordstr\"om geometry describes an empty space surrounding a charged black hole. The requisite is that the charge of the black hole be lesser than its mass. Because of this the geometry contains two horizons, an outer horizon and an inner horizon, but more important is the outer because this one give us the event horizon. The Reissner-Nordstr\"om metric is a static solution of the Einstein-Maxwell field equations, which corresponds to the gravitational field of a charged, non-rotating, spherically symmetric body of mass $M$ and charge $Q$\cite{r,n}. It is accepted that the laws of thermodynamics have close analogies in the physics of black holes. For example, the second law of thermodynamics is analogous to the second law of black hole dynamics, which implies that the surface of a black hole cannot decrease. In 1973 Bekenstein\cite{B} suggested that the area ${\cal A}$ of the event horizon of a black hole is a measure of its entropy: $S_{BH} = \frac{{\cal A}}{4 L^2_P}$. The Hawking temperature, entropy and mass of the black holes satisfy the first law of thermodynamics\cite{h,h1,h2}. However, all these important achievements are relative to the exterior description of black-holes. A description of the interior of black-holes would be desirable. The formulation of a consistent theory of quantum gravity is a very important in modern physics. There are some formulations that pretend provide a description of gravity at quantum level\cite{1bb,2bb,3bb,4bb,5bb}, which incorporate the geometry in different facets. There are other approaches, as the quantum geometrodynamics\cite{6bb,7bb}, and quantum geometry \cite{4bb,5bb}, which are more related to the idea of Einstein that all physical reality can be geometrized. Under certain conditions, the boundary conditions must be considered in the variation of the action\cite{Y}. In the event that a manifold has a boundary $\partial{\cal{M}}$, the action should be supplemented by a boundary term, in order for it to be
well defined\cite{GHa}. However, there is another way\cite{1m,2m} to include the flux that cross a 3D-hypersurface that
encloses a physical source without the inclusion of another term in the Einstein-Hilbert (EH) action. In this work we shall use Relativitic Quantum Geometry (RQG), which was introduced a few years ago\cite{1m,2m}. In this formalism, $\sigma$ is a massless scalar field that drives a geometrical displacement from a Riemannian manifold to an extended one, that leaves invariant the Einstein-Hilbert (EH) action after considering nonzero boundary terms to minimize the action. Back-reaction effects, that are responsible for the scalar fluctuations of space-time, obeys a dynamics on a background curved space-time characterized by a semi-Riemann manifold. In this last framework have been studied several topics. Some of them are inflationary back reaction effects\cite{8bb}, charged and electromagnetic fields \cite{9bb}, geometric back-reaction in pre-inflationary scenarios \cite{10bb,edg} and to study the initial state of the universe\cite{des}, among others.

The work is organized as follows: in Sect. II we revisit the boundary conditions in when we variate the Einstein-Hilbert (EH) action and the study
of the flux trough a closed 3D hypersurface when it is integrable. The formalism is revised and extended for nonconservative physical systems. In Sect. III we study these back-reaction effects in the interior of a Reissner-Nordstr\"om black hole (RNBH) as an isolated system. We define the mass and charge discretized labels also as the radius, and the energy, so that one can define discrete labels of entropy and the Bekenstein-Hawking temperature: $T_{BH}$. Finally, in Sect. IV we develop some final comments and conclusions.

\section{Boundary conditions in the minimum action principle and back-reaction effects}\label{2}

We consider the Einstein-Hilbert action ${\cal I}$, which describes gravitation and matter
\begin{equation}\label{act}
{\cal I} =\int_V d^4x \,\sqrt{-g} \left[ \frac{R}{2\kappa} + {\cal L}_m\right],
\end{equation}
where $\kappa = 8 \pi G$, $g$ is the determinant of the metric tensor $g_{\mu\nu}$, $R=g^{\mu\nu} R_{\mu\nu}$ is the scalar curvature,
$R^{\alpha}_{\mu\nu\alpha}=R_{\mu\nu}$ is
the covariant Ricci tensor and ${\cal L}_m$ is an arbitrary Lagrangian density which describes the physical fields.

The variation of the action is
\begin{equation}\label{delta}
\delta {\cal I} = \int d^4 x \sqrt{-g} \left[ \delta g^{\alpha\beta} \left( G_{\alpha\beta} + \kappa T_{\alpha\beta}\right)
+ g^{\alpha\beta} \delta R_{\alpha\beta} \right],
\end{equation}
with $g^{\alpha\beta} \delta R_{\alpha\beta} =\nabla_{\alpha}
\delta W^{\alpha}$, where  $\delta W^{\alpha}=
\delta\Gamma^{\epsilon}_{\beta\epsilon}{g}^{\beta\alpha}-\delta \Gamma^{\alpha}_{\beta\gamma} {g}^{\beta\gamma}$\cite{HE}. As in previous works\cite{1m,2m}, our
strategy will be to preserve the Einstein-Hilbert action as in
(\ref{act}) and see what are the consequences of do it. To set
$\delta {\cal I}=0$ in (\ref{delta}), we must consider the condition: $
G_{\alpha\beta} + \kappa T_{\alpha\beta} = \lambda(x)\,
g_{\alpha\beta}$, where $\lambda$ is the cosmological parameter. Additionally, we must require that
$g^{\alpha\beta} \delta R_{\alpha\beta} =\nabla_{\alpha}
\delta W^{\alpha} = \delta\Phi$, so that we obtain
the constraint $ \delta\Phi =-\lambda(x)\,g^{\alpha\beta}\,\delta g_{\alpha\beta} $ and therefore $\delta R_{\alpha\beta}=-\lambda(x)\, \delta g_{\alpha\beta}$. Here, we have used: $g^{\alpha\beta} \delta g_{\alpha\beta}=-g_{\alpha\beta} \delta g^{\alpha\beta}$. This means that exists a family of Einstein tensor transformations
\begin{equation}\label{ein}
\bar{G}_{\alpha\beta} = {G}_{\alpha\beta} - \lambda(x)\, g_{\alpha\beta},
\end{equation}
that leaves invariant the action:
\begin{eqnarray}
\bar{{G}}_{\alpha\beta}& = &- \kappa\,{{T}}_{\alpha\beta}.\label{tran1} \\
\end{eqnarray}

In order to calculate $\delta R_{\alpha \beta}$, we shall use the extended Palatini identity\cite{pal}
\begin{equation}
\delta{R}^{\alpha}_{\beta\gamma\alpha}=\delta{R}_{\beta\gamma}= \left(\delta\Gamma^{\alpha}_{\beta\alpha} \right)_{| \gamma} - \left(\delta\Gamma^{\alpha}_{\beta\gamma} \right)_{| \alpha}.
\end{equation}
We consider a geometry characterized the Levi-Civita connections plus a displacement with respect to the Riemann manifold
\begin{equation}\label{ga}
\Gamma^{\alpha}_{\beta\gamma} = \left\{ \begin{array}{cc}  \alpha \, \\ \beta \, \gamma  \end{array} \right\}+ \delta\Gamma^{\alpha}_{\beta\gamma}.
\end{equation}
where $\delta\Gamma^{\alpha}_{\beta\gamma}=\epsilon\,\sigma^{\alpha}\,g_{\beta\gamma} $. Here, $\sigma^{\alpha}\equiv \partial^{\alpha}\sigma$ and the extended covariant derivative of the metric tensor is nonzero (non-zero non-metricity). The value $\epsilon=1/3$ in $\delta\Gamma^{\alpha}_{\beta\gamma}$ assures the integrability of boundary terms. This means that in this case the boundary terms, which are in general given by
\begin{equation}
\delta\Phi=g^{\alpha\beta} \delta R_{\alpha\beta} =
\left[\delta W^{\alpha}\right]_{|\alpha} - \left(g^{\alpha\epsilon}\right)_{|\epsilon}  \,\delta\Gamma^{\beta}_{\alpha\beta} +
\left(g^{\alpha\beta}\right)_{|\epsilon}  \,\delta\Gamma^{\epsilon}_{\alpha\beta},
\end{equation}
take the particular form
\begin{equation}\label{gauge}
{g}^{\alpha \beta} \delta R_{\alpha \beta}-\delta \Phi =
\left[\delta W^{\alpha}\right]_{|\alpha} - \left(g^{\alpha\beta}\right)_{|\beta}  \,\delta\Gamma^{\epsilon}_{\alpha\epsilon} +
\left(g^{\epsilon\nu}\right)_{|\alpha}  \,\delta\Gamma^{\alpha}_{\epsilon\nu}= \nabla_{\alpha}\delta W^{\alpha}=-\nabla_{\alpha}\,\sigma^{\alpha}\equiv -\Box \sigma=0,
\end{equation}
which guarantees the integrability of the flux $\delta \Phi$ on the background metric. Notice that eq. (\ref{gauge}) is true only in the case with $b=1/3$, where the flux can be written in terms of a 4-divergence for $\delta W^{\alpha}$ defined in terms of covariant derivatives in the Riemann manifold. To calculate the flux $\delta\Phi$, we must know  $\delta R_{\alpha \beta}$. The variation of the Ricci tensor on the extended manifold is
\begin{equation}\label{VariacionRicciWeyl}
\delta R_{\alpha \beta}  = \left( \delta \Gamma^{\epsilon}_{\alpha \epsilon} \right)_{|\beta} - ( \delta \Gamma^{\epsilon}_{\alpha \beta})_{|\epsilon} \\
 = \frac{1}{3} \left[ \nabla_{\beta} \sigma_{\alpha} + \frac{1}{3} \left( \sigma_{\alpha} \sigma_{\beta} + \sigma_{\beta} \sigma_{\alpha} \right) - {g}_{\alpha \beta} \left( \nabla_{\epsilon} \sigma^{\epsilon} + \frac{2}{3} \sigma_{\nu} \sigma^{\nu} \right) \right].
\end{equation}
It is possible to write the left side of (\ref{gauge}) as
\begin{equation}\label{ein}
{g^{\alpha\beta}}\left\{ \frac{1}{3}\left[ \nabla_{\beta} \sigma_{\alpha} + \frac{1}{3} \left( \sigma_{\alpha} \sigma_{\beta} + \sigma_{\beta} \sigma_{\alpha} \right) - {g}_{\alpha \beta} \left( \nabla_{\epsilon} \sigma^{\epsilon} + \frac{2}{3} \sigma_{\nu} \sigma^{\nu} \right) \right]+\lambda(x)\,\delta g_{\alpha\beta}\right\} =0,
 \end{equation}
where the last term in (\ref{ein}) is due to the flux that cross the closed 3D-hypersurface: $\delta \Phi = -\lambda(x)\,g^{\alpha\beta}\,\delta g_{\alpha\beta}$. The variation of the metric tensor on the extended manifold will be also nonzero. The fact that $\delta g_{\beta\alpha}\neq 0$ on the extended manifold, means that the geometry described on this manifold is elastic in the sense that the norm of the tensors is not conserved. However, on the Riemann manifold this does not happen, because the non-metricity is null: $\Delta g_{\alpha\beta}=0$\footnote{In what follows we shall denote with a $\Delta$ variations on the Riemann manifold, and with a $\delta$ variations on an extended manifold.}
\begin{equation}\label{gab}
\delta g_{\alpha\beta} = g_{\alpha\beta|\gamma} \,dx^{\gamma} = -\frac{1}{3}\,\left[\sigma_{\beta} g_{\alpha\gamma} +\sigma_{\alpha} g_{\beta\gamma}
\right]\,dx^{\gamma}.
\end{equation}
 In general,
that makes the theory more powerful and able to describe non-conservative systems. Furthermore, the background space-time can be described from a quantum one:
\footnote{We can define the operator
\begin{displaymath}
\check{\delta x}^{\alpha}(t,\vec{x}) = \frac{1}{(2\pi)^{3/2}} \int d^3 k \, \check{e}^{\alpha} \left[ b_k \, \check{x}_k(t,\vec{x}) + b^{\dagger}_k \, \check{x}^*_k(t,\vec{x})\right],
\end{displaymath}
such that $b^{\dagger}_k$ and $b_k$ are the creation and destruction operators of space-time, such that $\left< B \left| \left[b_k,b^{\dagger}_{k'}\right]\right| B  \right> = \delta^{(3)}(\vec{k}-\vec{k'})$ and $\check{e}^{\alpha}=\epsilon^{\alpha}_{\,\,\,\,\beta\gamma\delta} \check{e}^{\beta} \check{e}^{\gamma}\check{e}^{\delta}$. }
\begin{equation}
dx^{\alpha} \left. | B \right> =  {U}^{\alpha} dS \left. | B \right>= \delta\check{x}^{\alpha} (x^{\beta}) \left. | B \right> ,
\end{equation}
is the eigenvalue that results when we apply the operator $ \delta\check{x}^{\alpha} (x^{\beta}) $ on a background quantum state $ \left. | B \right> $, defined on the Riemann manifold, and $g_{\alpha\beta|\gamma}$ denotes the covariant derivative on the extended manifold described by the connections (\ref{ga}). This quantum state can be represented in a ordinary Fock space. Notice that the equation (\ref{tran1}) describes the background (classical equations) with the flux that cross the 3D-gaussian hypersurface, which we shall consider that in this work is of quantum nature. Furthermore, ${U}^{\alpha}$ are the components of the Riemann
velocities. The line element is given by
\begin{equation}
dS^2 \, \delta_{BB'}=\left( {U}_{\alpha} {U}^{\alpha}\right) dS^2\, \delta_{BB'} = \left< B \left|  \delta\check{x}_{\alpha} \delta\check{x}^{\alpha}\right| B'  \right>.
\end{equation}
From the action's point of view, the scalar field $\sigma(x^{\alpha})$ drives a geometrical displacement from a Riemann manifold to the extended manifold, that leaves invariant the Einstein-Hilbert action
\begin{eqnarray}
{\cal I} &=& \int d^4 x\, {\sqrt{-g}}\, \left[\frac{R}{2\kappa} + {{\cal L}}\right] \nonumber \\
&=& \int d^4 x\, \left[\sqrt{-{g}} e^{-(2/3)\sigma}\right]\,
\left\{\left[\frac{{R}}{2\kappa} + {{\cal L}}\right]\,e^{(2/3)\sigma}\right\}. \label{aact}
\end{eqnarray}
Here ${R}$ is the Riemann scalar curvature, $\kappa= 8\pi G$, $G$ is the gravitational constant and ${\cal L}$ is the matter lagrangian density on the background Riemann
manifold. If we require that $\delta {\cal I} =0$, we obtain
\begin{equation}\label{gau}
-\frac{1}{V} \delta V = \frac{\delta \left[\frac{{R}}{2\kappa} + {{\cal L}}\right]}{\left[\frac{{R}}{2\kappa} + {{\cal L}}\right]}
= (2/3) \,\delta\sigma=\frac{1}{\lambda(x)}\,\delta\Phi,
\end{equation}
where $\delta\sigma = \sigma_{\mu} dx^{\mu}$ is an exact differential and ${ V}=\sqrt{-{ g}}$ is the volume of the Riemann manifold. Additionally, we shall require that $\sigma_{\alpha}\sigma^{\alpha}=0$, in order to (\ref{ein}) to be fulfilled. Of course, all the variations are in the extended manifold, and assure us gauge invariance because $\delta {\cal I} =0$.

\section{Reissner-Nordstr\"om black hole (RNBH) as an isolated system}

In this work we shall consider the case where the flux is null: $\delta\Phi=-\lambda(x)\,g^{\alpha\beta}\,\delta g_{\alpha\beta}=0$ (with $\delta g_{\alpha\beta}\neq 0$), and $\lambda(x)\equiv \lambda_0$. The first condition means that the physical system enclosed by the 3D-hypersurface is isolated, and the second one means that $\nabla_{\alpha} {G}^{\alpha\beta}=\nabla_{\alpha} \bar{G}^{\alpha\beta}=0$, so that there are no external sources (we must remember that the covariant derivative on the Riemann manifold is null: $\nabla_{\alpha} g_{\beta\epsilon}=0$). In this work we shall consider a description of space-time back reaction effects in the interior of a RNBH, which are described by the scalar field $\sigma$. In this framework back-reaction will describe the geometric response to the mass and charge distributions inside the isolated BH, because we are not considered a punctual charged mass in the center of the BH, but yes a spherically symmetric distribution.  Furthermore, outside the RNBH there is a physical vacuum and its mass and charge of the BH are considered as constant, there exterior does not interchange energy with the interior. Of course, this is an ideal system, but it is the started point to make a more realistic description.

\subsection{Back-reaction in the interior of a RNBH}\label{3}

We consider an isolated RNBH, such that $\delta\Phi=0$, and therefore the $\lambda(x)\equiv \lambda_0$. In this case the scalar field $\sigma$ is determined by $\Box \sigma=0$. Furthermore, notice that $\delta\Phi=0$, implies from (\ref{gau}), that $\frac{1}{V} \delta V=0$, so that the volume of the manifold remains constant, or $\sigma_{\mu}\,dx^{\mu}=0$. In spherical coordinates $(t,r,\theta,\phi)$, the line element that describes RNBH is
\begin{equation}\label{met}
dS^2 = -f(r) \; dt^2 + f(r)^{-1} \; dr^2 + r^2 \; d\Omega^2,
\end{equation}
where $d\Omega^2 = d\theta^2+\sin^2(\theta) \,d\phi^2$ and $f(r)=1-\frac{2M}{r}+\frac{Q^2}{r^2}$, such that a RNBH has two horizons; the event horizon and the Cauchy horizon: $r_{\pm}=M\pm\sqrt{M^2-Q^2}$.  Black holes with $Q^2>M^2$ can not exist in nature, because if the charge is greater than the mass there can be no physical event horizon. Notice that $f(r)<0$ in the range $r_-<r<r_+$. This is
the range in which we are aimed to study.

We must solve the equation
\begin{equation}
\Box \sigma = 0,
\end{equation}
in the range $r_-<r<r_+$. Here, $\Box$-differential operator is defined on the background metric described by the Levi-Civita symbols: $\Box\equiv \nabla_{\alpha}\nabla^{\alpha}$. As we are dealing with an isolated (and closed) system, it is expected that the charged mass distribution to be described by a finite number of degrees of freedom. We propose the change of variables: $u=\frac{1}{2}+\frac{r-M}{2\sqrt{M^2-Q^2}}$. Because we are dealing with spherical coordinates we can write the Fourier superposition for the field $\sigma(t,r,\theta,\phi)$
\begin{equation}
\sigma(t,r,\theta,\phi)= \sum_{n=0}^{N-1}\,\sigma_n(t,r,\theta,\phi),
\end{equation}
where
\begin{equation}
\sigma_n(t,r,\theta,\phi) = \sum_{l\geq L_{-}}^{L_{+}}\,\sum_{m =-l}^{l} \left[A_{nlm}\,\sigma_{nlm}(t,r,\theta,\phi)
+ A^{\dagger}_{nlm}\,\sigma^*_{nlm}(t,r,\theta,\phi)\right].
\end{equation}
The modes $\sigma_{nlm}(t,r,\theta,\phi)$, are:
\begin{equation}
\sigma_{nlm}(t,r,\theta,\phi) = \left(\frac{E_n}{\hbar}\right)^2 \, R_{nl}(r)\, Y_{lm}(\theta,\phi)\, T_n(t),
\end{equation}
where the functions $Y_{lm}(\theta,\phi)$ are the usual spherical harmonics. After making the substitution (a Wick rotation \cite{samuel}) $\tau=it$ on the metric (\ref{met}), we obtain that the temporal solution is
\begin{equation}
T_n(\tau)=B_1e^{i\frac{E_n}{\hbar}\tau}+B_2e^{-i\frac{E_n}{\hbar}\tau}.
\end{equation}
Furthermore, the radial solution is
\begin{eqnarray}\label{rad}
R_{nl}(u)=C_1(u-1)^{\frac{\gamma_n}{2}}u^{\frac{\beta_n}{2}}e^{\frac{\alpha_n}{2}u}{\cal H_C}(\alpha_n,\beta_n,\gamma_n,\delta_{n},\eta_{nl},u)
+ C_2(u-1)^{\frac{\gamma_n}{2}}u^{-\frac{\beta_n}{2}}e^{\frac{\alpha_n}{2}u}{\cal H_C}(\alpha_n,-\beta_n,\gamma_n,\delta_{n},\eta_{nl},u),
\end{eqnarray}
where the coefficients are:
\begin{subequations}
\begin{equation}
\alpha_n=4\;\sqrt{M^2-Q^2}\;\frac{E_n}{\hbar},
\end{equation}
\begin{equation}
\beta_n=\frac{\left[M^3(M-\sqrt{M^2-Q^2})-4MQ^2(2M-\sqrt{M^2-Q^2})+Q^4\right]^\frac{1}{2}}{(M^2-Q^2)^\frac{1}{2}}\frac{E_n}{\hbar},
\end{equation}
\begin{equation}
\gamma_n=\frac{\left[M^3(M+\sqrt{M^2-Q^2})-4MQ^2(2M+\sqrt{M^2-Q^2})+Q^4\right]^\frac{1}{2}}{(M^2-Q^2)^\frac{1}{2}}\frac{E_n}{\hbar},
\end{equation}
\begin{equation}
\delta_n=-8\sqrt{M^2-Q^2}M\frac{E_n^2}{\hbar^2},
\end{equation}
\begin{equation}
\eta_{nl}=\frac{8M(M^2-Q^2)\sqrt{M^2-Q^2}-4M^2(3Q^2-2M^2)-3Q^4}{2(M^2-Q^2)}\frac{E_n^2}{\hbar^2}-l(l+1).
\end{equation}
\end{subequations}
The values for $L_+$ and $L_-$ are determined by the values of $n$ and $-1<q=Q/M<1$, such that $\mu_{nl} \leq 0$\cite{Fiziev,esch}, where
\begin{equation}
\mu_{nl}=\frac{\alpha_n-\beta_n-\gamma_n+\alpha_n\beta_n-\beta_n\gamma_n}{2}-\eta_{nl}.
\end{equation}
They are defined by the solutions of the following quadratic equation:
\begin{align}
L_\pm&(-L_\pm+1)= \frac{1}{\left[q^2(q+4)(q-4)+4(f_2-f_1)\sqrt{1-q^2}-(f_1\,f_2+16)\right]} \nonumber \\
\times &\left[\left(8(1-q^2)^{\frac{3}{2}}+(f_2-4)f_1+(3q^2+4\,f_1-12)q^2-8\right)\,n^2 \right.\nonumber \\ &-\left(2(f_1-f_2-4(1-q^2))\sqrt{1-q^2}+6\,f_1(1-q^2)-(f_1-2)f_2-(7\,q^2+2\,f_2+16)q^2+8\right)\,n \nonumber \\
&-\left. 2\sqrt{1-q^2}\left((1+\sqrt{1-q^2})f_1)(1-\sqrt{1-q^2})f_2\right)+4q^2(q^2+1)\right],
\end{align}
where we have done the substitutions
\begin{equation}
f_1(q)=\sqrt{q^2(q^2-8)+4(q^2-2)\sqrt{1-q^2}+8}, \qquad f_2(q)=\sqrt{4\sqrt{1-q^2}\left(2(1+\sqrt{1-q^2}-1\right)+q^4}.
\end{equation}
In order to avoid divergent solutions, we must set $C_1=0$ in (\ref{rad}). Furthermore, following \cite{Fiziev}, we must impose the so called $\delta_{n}$ and $\Delta_{N+1}$ conditions, respectively
\begin{subequations}
\begin{equation}\label{dcond}
\frac{\delta_n}{\alpha_n}+\frac{\beta_n+\gamma_n}{2}+1=-n,
\end{equation}
\begin{equation}
\Delta_{N+1}=0,
\end{equation}
\end{subequations}
where $n$ is a positive integer. Once provided the fulfilment of the above two $\delta$-conditions, we obtain that the confluent Heun solutions is reduced to a polynomial of degree N\cite{Fiziev}. The condition (\ref{dcond}) for each particular solution of the radial equation, gives us the following expression for the energy levels $E_{n}$:
\begin{equation}\label{elevels}
E_{n}\,r_{*,n}=\frac{\hbar}{2},
\end{equation}
where the discrete radius values, are given by the expression
\begin{subequations}
\begin{eqnarray}
r_{*,n}=&&\frac{r_{+,n}+r_{-,n}}{2}+\frac{\left[8(r_{+,n}-r_{-,n})[(r_{+,n}+r_{-,n})^2+(r_{+,n}-r_{-,n})^2]r_{+,n}+[(r_{+,n}+r_{-,n})^2
-(r_{+,n}-r_{-,n})^2]^2\right]^\frac{1}{2}}{8(r_{+,n}+r_{-,n})} \nonumber \\ \label{pm}
&\pm& \frac{\left[8(r_{+,n}-r_{-,n})[(r_{+,n}+r_{-,n})^2+(r_{+,n}-r_{-,n})^2]r_{-,n}+[(r_{+,n}+r_{-,n})^2
-(r_{+,n}-r_{-,n})^2]^2\right]^\frac{1}{2}}{8(r_{+,n}+r_{-,n})},
\end{eqnarray}
with
\begin{equation}
r_{+,n}=M_{n}+\sqrt{M_n^2-Q_n^2}, \; r_{-,n}=M_{n}-\sqrt{M_n^2-Q_n^2}.
\end{equation}
Here, it is very important to notice that $M_n$ and $Q_n$ are discrete values for the mass and the charge, given by
\begin{equation}
M_{n}=\frac{M}{(n+1)}, \; Q_{n}=\frac{Q}{(n+1)}.
\end{equation}
\end{subequations}
The expression (\ref{elevels}) is very important because it tells us that the uncertainly principle is fulfilled for each energy level. The signature $\pm$ in (\ref{pm}) corresponds to consider respectively ${\cal H_C}(\alpha_n,\beta_n,\gamma_n,\delta_{n},\eta_{nl},u)$ or ${\cal H_C}(\alpha_n,-\beta_n,\gamma_n,\delta_{n},\eta_{nl},u)$. If $Q=0$, hence $r_{-,n}=0$ and therefore $r_{+,n}=2M_{n}=r_{sch,n}=r_{*,n}$. In this case, we recover the results obtained in \cite{esch}. Furthermore, if $M=Q$, hence we obtain that $r_{*,n}$ is only well defined for ${\cal H_C}(\alpha_n,-\beta_n,\gamma_n,\delta_{n},\eta_{nl},u)$, and in this case we have $r_{+,n}=r_{-,n}=r_{*,n}=M_n\equiv Q_{n}$.

\subsubsection{Charge and mass discretization}

If we suppose that exist a condition $M>Q$ for avoid naked singularities, we obtain a condition for the possible values of $n$:
\begin{equation}
n \leq \frac{M}{Q}-1 \leq N-1, \label{mq}
\end{equation}
where $N$ is the lower integer number that verify $\frac{M}{Q} \leq N$. This is a very important result because say us the maximum number of states $N$ is given by the ratio between the classical mass $M$ and the classical charge $Q$.

We calculate the difference between two consecutive $Q_{n}$ states, named $\Delta Q_{n}$:
\begin{equation}
\Delta Q_n = Q_n - Q_{n+1} = \frac{Q}{(n+1)(n+2)}
\end{equation}
The summation of $\Delta Q_n$ on the possible range of $n$ values, when it tends to infinity give us the classic value of charge $Q$:
\begin{equation}
\lim_{N\rightarrow \infty} \sum\limits_{n=0}^{N-1}\Delta Q_{n} = Q.
\end{equation}
We can adopt an analogous procedure with the difference between two consecutive $M_{n}$ values\cite{esch}. In this case using $r_{+,n}$ and $r_{-,n}$, we obtain:
\begin{subequations}
\begin{equation}
\Delta M_n = \frac{\left(\Delta r_{+,n} + \Delta r_{-,n}\right)}{2} = \frac{M}{(n+1)(n+2)},
\end{equation}
where we variations of $r_{\pm,n}$ radius, are
\begin{equation}
\Delta r_{\pm,n} = r_{\pm,n} - r_{\pm,n+1}.
\end{equation}
\end{subequations}
Finally, when the upper cut $N$ tends to infinity, we obtain that the summation of $\Delta M_n$ on the possible range of $n$ values is the classical $M$-mass
\begin{equation}
\lim_{N\rightarrow \infty} \sum\limits_{n=0}^{N-1}\Delta M_{n} = M,
\end{equation}
in agreement with the result obtained in\cite{esch}.

\subsubsection{Quantum thermodynamics}

The discrete values of $r_{+,n}$ allow us to also discretize the exterior area $A_{+,n}$ and its corresponding entropy $S_{+,n}$. We can determine the difference between two consecutive exterior area states: $\Delta A_{+,n}>0$:
\begin{equation}
\Delta A_{+,n} = 4 \pi \left(r_{+,n}^{2} - r_{+,n+1}^{2}\right) = 4 \pi r_+^{2} \left( \frac{2n+3}{(n+1)^{2}(n+2)^{2}} \right).
\end{equation}
Furthermore, we can calculate the total contribution of the exterior area, as the sum over the difference between consecutive $nth$-exterior areas:
\begin{equation}
\sum\limits_{n=0}^{N-1}\Delta A_{+,n} = 4 \pi r_+^{2} \left( 1-\frac{1}{(N+1)^{2}} \right) ,
\end{equation}
that depends on the number of states, $N$. In the same way, we obtain the entropy from the sum over the difference between consecutive $nth$-entropies:
\begin{equation}
\sum\limits_{n=0}^{N-1}\Delta S_{n} =   \frac{1}{4}\,\sum\limits_{n=0}^{N-1}\Delta A_{+,n} = \pi r_+^2 \left( 1-\frac{1}{(N+1)^{2}} \right).
\end{equation}
Finally, if we calculate the temperature $T_N$ in the context of the first law of B-H mechanics:
\begin{equation}
T_{N}=\frac{\sum\limits_{n=0}^{N-1} \Delta M_{n}}{\sum\limits_{n=0}^{N-1} \Delta S_{n}}=\frac{\left(r_+-r_-\right)}{2 \pi r_+^2}
\frac{1}{\left( 1 + \frac{1}{N+1} \right) },
\end{equation}
we obtain that $T_{BH} \leq T_{N} < 2\,T_{BH}$\cite{esch}, for different values of $N$. However, this is not a true temperature, because in the interior of the black hole the states are entangled, so that $T_N$ must be interpreted as a {\it latent temperature}. It is interesting to notice that, if we have a single state, i.e., if $N=1$, we recover the Bekenstein-Hawking temperature for Reissner-Nordstr\"om, which could be related to a true temperature because this state is associated to an energy level which is at the exterior limit radius: $r_{+,0}=M_0+\sqrt{M^2-Q^2}=r_+$. When the number of states is very large ($N\gg1$), we obtain the up cut for $T_N$:
\begin{equation}
\lim_{N\rightarrow \infty} T_N= \lim_{N\rightarrow \infty} \frac{\left(r_+-r_-\right)}{2 \pi r_+^2} \frac{1}{\left( 1 + \frac{1}{N+1} \right) } \rightarrow \frac{(r_+-r_-)}{2 \pi r_+^2} =2\, T_{BH}.
\end{equation}
This means that the latent quantum temperature will be in the range: $T_{BH} \leq T_{N} < 2\,T_{BH}$.

\section{Final comments}\label{4}

We have used the formalism of Relativistic Quantum Geometry to study the quantum structure of space-time in the interior of a Reissner-Nordstr\"om Black-Hole, without angular moment. The radial solution can be described as a local solution around $r_+=M+\sqrt{M^2-Q^2}$, so that it can be expressed as a finite superposition of polynomials that describe the energy levels of space-time. The maximum number of states is determined in expression (\ref{mq}); {\em the maximum number of states $N$ is given by the ratio between the classical mass $M$ and the classical charge $Q$}. Due to this result, we have verified that the uncertainly principle, written as $E_n\, r_{sh,n}=\hbar/2$, is fulfilled on each energy level of the RNBH. This provide us with the possibility to develop a quantum thermodynamic, where energy, mass, charge, temperature and entropy are dependent with the number of states in the RNBH. Finally, it should be highlighted that the classical RNBH entropy is defined as the limit case where the number of states is very large, and the RNBH temperature is obtained for a black-hole with an isolated state. The quantum temperature depends on the number of states in the interior of the RNBH. It is always in  the range: $T_{BH} \leq T_{N} < 2\,T_{BH}$, and must be considered as a latent temperature because the quantum coherence of the system\cite{ch2,ch1,ch,da}. This can be assured because we have supposed a null flux $\delta \Phi=0$, so that the RNBH constitutes an isolated physical system. Of course, this is an idealised analysis of the system, but this is a started point to study what should happen in the interior of the BH when there is an interchange of energy with a non vacuum exterior. Anyway, the topic here studied is very important because help us to understand the quantum nature in the interior of the RNBH, and because there are "levels" of quantum space-time, and therefore of charged mass,
which are discretized analogously to that of the electronic energy levels in an atom.

\section*{Acknowledgements}

\noindent The authors acknowledge CONICET, Argentina (PIP 11220150100072CO) and UNMdP (EXA955/20) for financial support.

\end{document}